\documentstyle[prl,aps,epsf]{revtex}
\draft
\begin{document}
\twocolumn[\hsize\textwidth\columnwidth\hsize\csname @twocolumnfalse\endcsname

\title{Individual and Multi Vortex Pinning in Systems with Periodic Pinning Arrays}

\author{C.~Reichhardt}
\address{CNLS and Applied Theoretical and Computational Physics Division,
Los Alamos National Laboratory, Los Alamos, NM 87545} 

\author{G.T.~Zim{\' a}nyi, R.T.~Scalettar}
\address{Department of Physics, University of California, Davis, California
95616.}

\author{A.~Hoffmann}
\address{Los Alamos National Laboratory, Los Alamos, NM 87545}

\author{Ivan K.~Schuller}
\address{University of California-San Diego, Department of Physics 0319,
La Jolla, California 92093-0319} 

\date{\today}
\maketitle
\begin{abstract}
We examine multi and individual vortex pinning in thin superconductors with 
periodic pinning arrays. For multi-vortex pinning we observe peaks in 
the critical current of equal magnitude at every matching field, while for
individual vortex pinning we observe a sharp drop in the critical current
after the first matching field in agreement with experiments.
We examine 
the scaling of the critical current at commensurate and incommensurate
fields for varied pinning strength and show that the
depinning force at the incommensurate fields decreases faster than 
at the commensurate fields. 
\end{abstract}

\vskip2pc]
\narrowtext

Vortex matter in superconductors with periodic pinning arrays has been studied
for several years now starting 
with the pioneering work of Fiory {\it et al.} \cite{Fiory}. 
Recently considerable renewed interest in this system has come about due to
advancements in nano-lithographic techniques in which arrays of 
holes \cite{Metlushko,Fractional,Welp,Bezryadin,Field}, 
defects \cite{Harada} or
magnetic dots \cite{Schuller,Velez,Lange,Fasano,Jaccard,Terentiev} 
can be created where various parameters such as the pinning
strength, size, and periodicity can be carefully controlled. These 
experiments have observed interesting commensurability effects where 
the critical current shows a maximum 
and magnetoresistance a minimum when the number of vortices equals an
integer multiple of the number of pinning sites. 
In addition peaks in the critical 
current have also been observed at 
some fractional matching fields \cite{Fractional}.
Many applications of superconductors require high critical currents
and the nano-engineered periodic pinning systems may be able to provide 
optimal pinning.   
A key question that needs to be understood in order for optimal
pinning arrays to be constructed  
is where and how the vortices are arranged at the 
varied matching fields.  In particular, it is not clear whether there 
are multiple vortices located
at individual pining sites above the first matching field or whether only one 
vortex sits at an individual pinning site, nor has it been shown 
how this ordering affects the
observed commensuration effects.

In magnetization experiments in 
anti-dot lattices \cite{Metlushko} 
it was observed that the critical current dropped abruptly
beyond a specific matching field which depends on temperature.
It was inferred that below this field
the pinning sites were capturing multiple vortices that could be strongly 
pinned, and that beyond this field 
the additional vortices were located in the interstitial
regions and were not pinned directly by the pinning sites. 
On the other hand imaging experiments using Lorentz microscopy \cite{Harada} 
of vortices in periodic pinning arrays observed only one vortex being 
trapped per pinning site and showed
that at the integer matching fields the vortices
form highly ordered crystals with the crystal symmetry depending on 
the particular matching field. 
Subsequent simulations of vortices in periodic pinning
arrays where only one vortex can be trapped at a pinning site also produced
the same types of vortex crystals 
observed in the Lorenz microscopy experiments at 
the matching fields \cite{Commensurate}. 
In addition the simulations also showed that
peaks in the critical current occur at those matching fields where the 
interstitial vortices can form a highly ordered structure; however, some 
matching fields did not show a peak in the critical current, and 
the magnitude of the commensuration effect varied considerably at 
different matching fields. Recent experiments with small pinning sites
where it is expected
that only one vortex can be pinned at an individual pinning sites observed
commensuration peaks in agreement with these 
simulations \cite{Welp}.
There is also evidence from decoration experiments \cite{Bezryadin}
that multi-vortex pinning
at individual pinning sites in periodic pinning arrays 
occurs. 
Other recent  
imaging experiments with Hall probe arrays 
have also observed multi-vortices 
at individual pinning sites and show that above a certain field, additional
vortices sit in the interstitial regions \cite{Field}.
These experiments seem to indicate that
when multi-vortex pinning occurs the vortices in the pinning site merge
to form a macro-vortex.    
       
Another open question in vortex pinning in periodic arrays is why 
the commensuration effects are only pronounced near $T_{c}$. 
One proposal is that the bulk pinning increases at low $T$ and washes
out the effect of the periodic pins; however, samples with periodic 
pinning always show a much higher critical current than
comparison samples without the periodic pinning even for low
$T$, indicating that the periodic pinning is still effective at
low $T$. 

In this work we present results from simulations to 
elucidate the response of the critical current with  
multi-vortex vs individual vortex pinning. We model the 
multi vortices as individual vortices with increased number of flux quanta.    
We find that with multi vortex pinning, peaks in the critical current 
of the same magnitude occur
at every matching field.
For individual vortex pinning the critical current drops off after the 
first matching field. 
These results agree well with experimental results for vortices interacting
with magnetic dots where with large dots multi-vortex pinning could be
expected.
We also examine the scaling of the depinning force 
at commensurate and incommensurate fields as a function of  
the pinning strength. For all pinning strengths examined we find that
depinning force scales linearly with pinning strength at commensurate fields.
For incommensurate fields the depinning force scales linearly for large 
pinning forces but shows a crossover to a faster than linear scaling with
smaller pinning strengths.  
This result may explain why commensurability 
effects are most pronounced for high temperatures where the pinning is
reduced. 
In addition we also examine the scaling of the critical current 
at the second matching field for multi-vortex pinning and individual pinning
and find that as the pinning strength is increased the critical current
saturates when individual vortex pinning is present. 

We model vortices in a 2D superconductor interacting with a square pinning
array. The overdamped equation of motion for a vortex $i$ is 
\begin{equation}
{\bf f}_{i} = \frac{d {\bf r}_{i}}{dt} = 
{\bf f}_{i}^{vv} + {\bf f}_{i}^{vp} + {\bf f}_{d} = \eta{\bf v}_{i}
\;
\end{equation}
Here, ${\bf f}_{i}^{v}$ is the force from the other vortices,
${\bf f}_{i}^{vp}$ is the force from the pinning, 
${\bf f}_{d}$ is the applied driving force corresponding to a 
Lorentz force from an applied current, and $\eta$ is the damping which
we set to unity.  
We model the vortex-vortex interaction potential as $U_{v} = \ln (r)$ which
is appropriate for thin film superconductors. We have also conducted 
simulations using the modified Bessel function 
interaction potential $K_{0}(r)$ which is appropriate for bulk superconductors.
The total force on vortex $i$ from the other vortices 
is $\sum_{j\neq i}^{N_{v}}\nabla_{i} U_{v}(r_{ij})$. We impose periodic
boundary conditions in the $x$ and $y$ directions. For the 
long-range logarithmic 
interaction we use an exact and fast converging sum \cite{Jensen}.
The pinning is modeled as attractive parabolic wells of radius $r_{p}$.
\begin{equation}
f_{i}^{vp} = (f_{p}/r_{p})({\bf r}_{i}-{\bf r}_{k}^{p}) \Theta(|{\bf r}_{i} 
- {\bf r}_{k}^{p}|/\lambda).
\end{equation}
Here, $\Theta$ is the Heaviside step function, ${\bf r}_{k}^{(p)}$ is the
location of pinning site $k$, $f_{p}$ is the maximum pinning force, 
and $\lambda$ is the penetration depth. 
The pinning is placed in a square array.    
The initial vortex positions are obtained from annealing from a high 
temperature where the vortices are in a molten state and gradually 
cooling to $T = 0$. After the vortex configurations are obtained the
critical depinning force is determined by applying a slowly increasing 
uniform force. We simulate two models.  In the first, the pinning 
sites are small and can capture only one vortex each,
so for fields greater than the first matching field additional
vortices are located between the pinning sites. In the second model,
multiple vortices can be captured by larger pinning sites, and
we assume
that the vortices in the pinning sites coalesce and form a single
multiple-quantized vortex. The multi-vortices interact with other vortices
as $n ln(r)$ where $n$ is the quantization of the vortex. 
Beyond the first matching field we start
the simulation by annealing
with the number of multiple quantized vortices already fixed so that
we do not actually model the merging of the vortices. The multi-quantized
vortices will still feel the same pinning force as the individual 
vortices.  This is a reasonable assumption 
if the core of the vortex is smaller than the pinning site, 
since the maximum pinning force 
is determined by the gradient of the potential energy of the 
pinning site, which should not be affected by the size of the vortex. 

In Fig.~1 we show the critical depinning force $f_{p}^{c}$ versus field
for the case where multiple vortex pinning occurs
(thick curve) and for the case of single 
vortex pinning (thin curve). For $B/B_{\phi} < 1.0$ the depinning force is   
the same for both pinning radii which is due to the fact that the maximum 
pinning force for the pinning sites are the same. A peak at 
$B/B_{\phi} = 1/2$ occurs which is due to the formation of an ordered state as
seen in previous simulations and experiments. There is a peak in both 
curves at the the first matching field where the vortex lattice has the 
same symmetry as the pinning lattice. 
For fields greater than 
$B/B_{\phi} = 1.0$ the curves deviate, with the single vortex pinning curve
dropping substantially while the multiple vortex pinning curve only decreases
by a small amount. The height of  the peak for the multi-vortex 
curve at 
$B/B_{\phi} = 2.0$ 
and the higher matching fields 
is the same as that for $B/B_{\phi} = 1.0$ since in our
model the multiple vortices experience the same pinning force 
as the singly quantized vortices. The multi-vortex curve 
also shows matching peaks at the fractional fields $n/2$.  
For the single vortex pinning curve the maximum 
critical depinning force at $B/B_{\phi} =  2.0$ is much lower than that 
for the multi-vortex case. The fractional matching peaks are also 
suppressed for the individual vortex pinning case.    

The vortex configurations where only one vortex is captured per pinning 
site are the same as those observed in 
previous experiments and simulations. 
The vortex configurations 
for the multi-vortex pinning case repeat for each matching field with 
a square overall vortex lattice at each matching field 
composed of multiply quantized vortices.     

In Fig.~2 we show the results of
transport experiments for Nb films with 
square periodic arrays of 
magnetic Ni dots for two different sized dots. Here the period 
is $a = 600 nm$ and the upper curve has a dot diameter of 
$d= 400-nm$ and the lower curve $d = 530-nm$ and $T = 0.98T_{c}$. 
Additional information about the fabrication of the arrays can be found in 
Ref[8]. The values of $\rho(H)$ for both sized dots is roughly the 
same for $B/B_{\phi} = 1.0$ with the $\rho$ minima values at 
$B/B_{\phi} = 1.0$ being the same. For $B/B_{\phi} > 1.0$ 
the $d = 400-nm$ pinning is greatly reduced as seen by the increase 
in $\rho(H)$;  however, minima still occur at the higher matching fields. 
In contrast the $\rho$ values for the $d = 530-nm$ show only a
minimal increase with $H$ and pronounced matching minima of 
approximately the same magnitude at each field are observed. 
This result agrees well
with the simulations, suggesting that for the larger dots,
multi-vortex pinning 
is occurring at each matching field, while for the smaller
dots, only one vortex is being trapped at each dot and additional
vortices are located in the interstitial regions.   

In Fig.~3 we show that commensurability effects
become more pronounced for weaker pinning. 
We consider   
multi-vortex pinning for $f_{p} = 0.8$ and $0.1$. Fig.~3 shows
$f_{p}^{c}/f_{p}$ for the two pinning strengths. At the matching fields
the depinning force equals the pinning force. At the incommensurate fields
the relative depinning force is much lower for the weaker pinning sample. 

In order to quantify the behavior observed in Fig.~3,  
in Fig.~4(a) we examine the depinning force
for the commensurate field $B/B_{\phi} = 1.0$ and
the incommensurate field $B/B_{\phi} = 0.64$
for a series of simulations with varied $f_{p}$. 
The depinning force for $B/B_{\phi} = 1.0$ 
decreases linearly with $f_{p}^c \propto f_{p}$ which is the individual
pinning regime. 
Since the vortex lattice is perfectly symmetric at the matching field, the
vortex-vortex interactions cancel so the depinning force is determined only
by the pinning energy. For $B/B_{\phi} = 0.64$, 
where vortex-vortex interactions should be relevant, we observed that for 
large $f_{p}$ the depinning force 
again decreases linearly with $f_{p}$; however, near $f_{p} = 0.2$ there is a
crossover to faster 
than linear decrease with $f_{p}^{c} \propto f_{p}^{1.42}$. 
We have found the same behavior for other incommensurate fields; 
thus, the difference in the critical current between the commensurate and the
incommensurate fields grows as $f_{p}$ is lowered. 
This behavior could account for the pronounced matching peaks only
near $T_{c}$ where the pinning is weak.
We were not able to 
go to weaker pinning where one might expect a crossover to a
collective pinning regime with $f_{p}^{c} \propto f_{p}^{2}$. 

To compare the effect of pinning strength on the critical current 
for individual versus multiple vortex pinning we have conducted a series
of simulations at $B/B_{\phi} = 2.0$ for varied $f_{p}$ for the 
two cases. In Fig.~4(b) we show that $f_{p}^{c}$ 
for the multi-vortex pinning case increases linearly with $f_{p}$. 
We have also found the same linear increase at the other matching
fields for the multi-vortex pinning case. The depinning in 
the multi-vortex case occurs in one stage with the entire lattice depinning 
at once. 
For the individual vortex pinning case there is an initial increase in 
$f_{p}^{c}$ for low values of $f_{p}$ followed by a saturation. 
This saturation can be understood by considering that the pinning 
of the interstitial vortices is caused by the repulsion of the vortices
at the pinning sites which is independent of the pinning strength. 
The saturation point also marks the onset of a two stage
depinning process where the interstitial vortices depin first
followed by the vortices at the pinning sites at a higher drive.
A two stage depinning transition at $B/B_{\phi} = 2.0$ has been observed
in experiments for widely spaced hole arrays \cite{Rossel}. 
Below the saturation point the entire lattice depins simultaneously. 

In conclusion we have compared multiple and individual vortex pinning 
in periodic pining arrays. We find that with multiple vortex pinning, 
peaks in the critical depinning force occur at every matching field with 
the same amplitude. For individual vortex pinning the depinning force
drops markedly for $B/B_{\phi} > 1.0$; however, peaks at the matching field
are still present. These results are in good agreement with transport
measurements on periodic magnetic dot arrays for different dot sizes with
the smaller dot systems showing results consistent with individual pinning
while the larger dot systems show results constant with multiple vortex
pinning. We also show that the commensuration effects are more 
pronounced for weak pinning, with the critical depinning force
decreasing linearly with decreasing pinning force while 
the  depinning force at incommensurate fields decreases 
faster than linear. For multi-vortex pinning the critical depinning force
scales linearly with pinning force at all matching fields while for 
individual pinning there is a saturation effect of the critical depinning
force.  

{\bf Acknowledgments} We thank C.J.~Olson for critical reading of this
manuscript, and S.~Bending, S.~Field, and V.~Metlushko for useful
discussions. This work 
was supported by NSF-DMR-9985978, 
CLC and CULAR (Los Alamos National Laboratory).

\begin{figure}
\center{
\epsfxsize=3.5in
\epsfbox{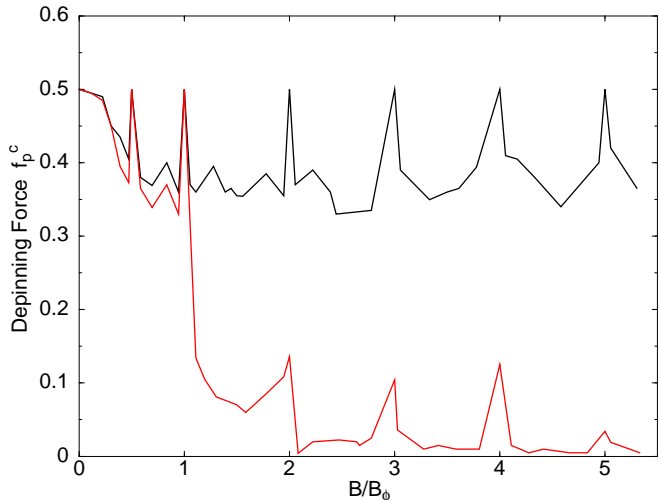}}
\caption{
Critical depinning force as a function of vortex density for 
samples with square arrays of pinning sites. 
The upper curve (thick lines) is the depinning system with
multiple vortex pinning for $B/B_{\phi} > 1.0$, $r_{p} = 0.5$ and
lower curve is the depinning line for a system when only one vortex
is captured at a pinning site.}
\label{fig1}
\end{figure} 

\begin{figure}
\center{
\epsfxsize=3.5in
\epsfbox{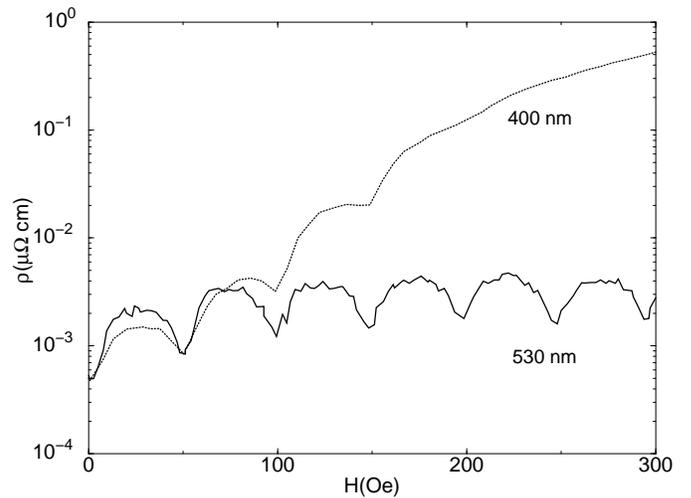}
}
\caption{
Experimental values of $\rho(H)$ for Nb samples with square 
arrays of Ni dots for two different dot diameters 
530 nm (solid line) and 400 nm (dashed line) at $T/T_{c} = 0.98$.}
\label{fig2}
\end{figure} 

\begin{figure}
\center{
\epsfxsize=3.5in
\epsfbox{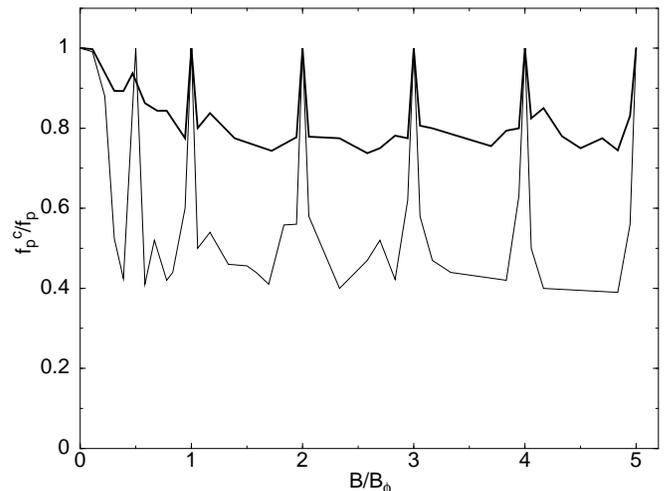}}
\caption{ 
The normalized depinning force vs $B/B_{\phi}$ for $f_{p} = 0.8$ upper curve
(thick lines) and $f_{p} = 0.1$ lower curve (thin line) obtained from
simulations. Multi-vortex pinning occurs for $B/B_{\phi} > 1.0$. 
Here the 
commensurability effects can be seen to be more pronounced for the
weaker pinning sample.}
\label{fig3}
\end{figure} 

\begin{figure}
\center{
\epsfxsize=3.5in
\epsfbox{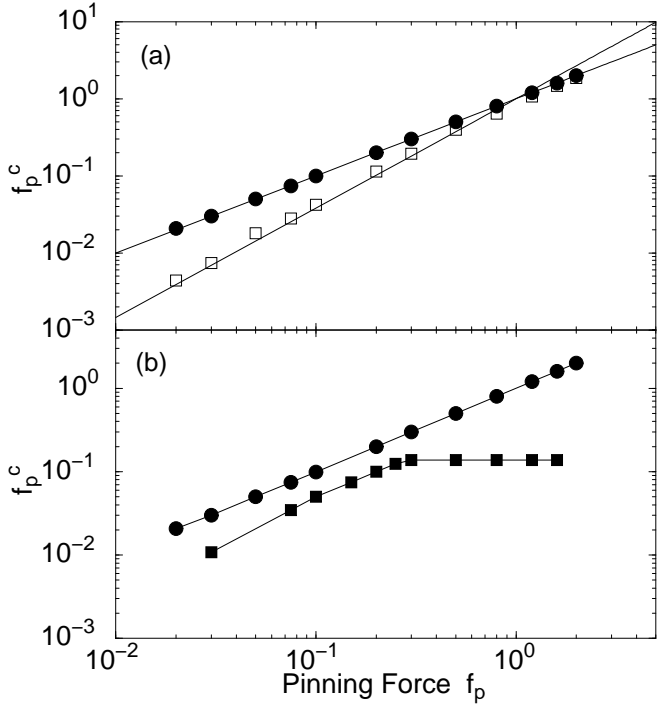}}
\caption{ 
(a) The depinning force 
vs $f_{p}$ for $B/B_{\phi} = 1.0$ (filled circles) and 
$B/B_{\phi} = 0.64$ (open circles). The depinning force for 
$B/B_{\phi}$ goes linearly with $f_{p}$ as shown with the fit while
the depinning force for $B/B_{\phi} = 0.64$ goes as $f_{p}^{1.42}$ as 
shown with the fit. For $f_{p} > 1.0$ the depinning force for 
$B/B_{\phi} = 0.64$ crosses over to a linear behavior. 
(b) The depinning force for $B/B_{\phi} = 2.0$ for multi-vortex
pinning (filled circles) and for single vortex pinning (filled squares). 
For the single vortex pinning there is a saturation in the depinning
force for $f_{p} > 0.3$ while the multi-vortex pinning case increases  
linearly with $f_{p}$.}
\label{fig4}
\end{figure}

\end{document}